\documentclass[12pt]{article}
\usepackage{epsfig}\parskip 5pt plus 1pt
\usepackage{bbm}
\usepackage{amsmath}
\usepackage{amssymb}
\usepackage{amsfonts}
\usepackage{mathrsfs}
\usepackage{graphicx}
\usepackage{color}
\newcommand{\nn}{\nonumber}

\newcommand{\dv}{\partial\hspace{-7pt}\slash}

\newcommand{\be}{\begin{equation}}
\newcommand{\ee}{\end{equation}}
\newcommand{\bea}{\begin{eqnarray}}
\newcommand{\eea}{\end{eqnarray}}

\begin{document}
\begin{titlepage}
\vspace*{-2cm}
\flushright{FTUAM 08-02\\ IFT-UAM/CSIC 08-12\\ LPT-Orsay 08-05\\ULB-TH/08-04\\MPP-2008-11}
\vskip 1.5cm
\begin{center}
{\Large\bf $\boldsymbol{\mu \rightarrow e \gamma}$ and $\boldsymbol{\tau \rightarrow l \gamma}$ decays in the fermion\\ \vspace{3mm}
triplet seesaw 
model}
\end{center}
\vskip 0.5cm
\begin{center}
{\large A.~Abada}$\,^a$~\footnote{asmaa.abada@th.u-psud.fr},
{\large C.~Biggio}$\,^b$~\footnote{biggio@mppmu.mpg.de},
{\large F.~Bonnet}$\,^a$~\footnote{florian.bonnet@th.u-psud.fr}\\
\vskip .1cm
{\large M.B.~Gavela}$\,^c$~\footnote{belen.gavela@uam.es} and 
{\large T.~Hambye}$\,^d$~\footnote{thambye@ulb.ac.be}\\
\vskip .7cm
$^a\,$ Laboratoire de Physique Th\'eorique  UMR 8627,\\
Universit\'e de Paris-Sud 11, Bat. 210, 91405 Orsay Cedex, France
\vskip .1cm
$^b\,$ Max-Planck-Institut f\"ur Physik,\\
80805 M\"unchen, Germany
\vskip .1cm
$^c\,$ Departamento de F\'\i sica Te\'orica and Instituto de F\'\i sica Te\'orica UAM/CSIC,
\\
Universidad Aut\'onoma de Madrid, 28049 Cantoblanco, Madrid, Spain
\vskip .1cm
$^d\,$ Service de Physique Th\'eorique,\\
Universit\'e Libre de Bruxelles, 1050 Brussels, Belgium
\end{center}
\vskip 0.5cm

\begin{abstract}
In the framework of the seesaw models with triplets of fermions, we
evaluate the decay rates of $\mu \rightarrow e \gamma$ and $\tau
\rightarrow l \gamma$ transitions. We show that although, due to
neutrino mass constraints, those rates are in general expected to be
well under the present experimental limits, this is not necessarily
always the case. Interestingly enough, the observation of one of those
decays in planned experiments would nevertheless contradict bounds
stemming from present experimental limits on the $\mu \rightarrow eee$
and $\tau \rightarrow 3 l$ decay rates, as well as from $\mu$ to $e$ 
conversion in atomic nuclei.  Such detection of radiative
decays would therefore imply that there exist sources of lepton
flavour violation not associated to triplet fermions.

\end{abstract}
\end{titlepage}
\setcounter{footnote}{0}
\vskip2truecm

\newpage
\section{Introduction}

The search for flavour changing rare leptonic decays, in particular
for  $\mu \rightarrow e \gamma$, $\tau \rightarrow \mu \gamma$ and
$\tau \rightarrow e \gamma$ decays, has been the object of intense
experimental investigations for decades~\cite{PDG}. With respect to
the present experimental upper limit, Br$(\mu \rightarrow e \gamma) <
1.2 \cdot 10^{-11}$~\cite{PDG}, Br$(\tau \rightarrow \mu \gamma) < 4.5
\cdot 10^{-8}$~\cite{Hayasaka:2007vc}, Br$(\tau \rightarrow e \gamma)
< 1.1 \cdot 10^{-7}$~\cite{PDG}, new experiments are expected to
improve in the near future their branching ratios by as much as three
orders of magnitudes for the first decay mode~\cite{meg} and by one or
two for the two others~\cite{newtau}.

The recent experimental evidence for neutrino masses has shown that
lepton flavour is violated in the neutrino sector and that,
consequently, in a model independent way, these decay rates are
predicted to be different from zero.  The actual predicted rate,
however, turns out to be highly model dependent.  There are three
basic models which can explain the neutrino masses at tree level, from
the exchange of heavy states, through the seesaw mechanism.  The above
rare decays have been studied at length in the framework of two of
these seesaw models, with right-handed neutrinos~\cite{muegammatype-I}
(type-I seesaw~\cite{Seesaw}) and with one or several Higgs
triplets~\cite{higgstriplets} (type-II seesaw~\cite{type-II}). In this
letter we perform the calculation of these decay rates in the
framework of the third seesaw model, with heavy triplets of fermions
(type-III seesaw~\cite{type-III}). This model has been studied in
detail, both from the theoretical and phenomenological point of view,
in Ref.~\cite{ABBGH}, where the result on these rare decays has
already been presented without the detailed calculation.
This letter also contains a determination of the constraint that $\mu$ to $e$ 
conversion in atomic nuclei implies on the type-III seesaw model.

\section{The type-III seesaw Lagrangian}

The type-III seesaw model consists in the addition to the standard
model of SU(2) triplets of fermions with zero hypercharge, $\Sigma$.
In this model at least two such triplets are necessary in order to
have two non-vanishing neutrino masses.  A non-vanishing $l_1
\rightarrow l_2 \gamma$ rate can nevertheless be induced already with
only one fermionic triplet. In the following, we will not specify the
number of triplets so that our calculation is valid for any number of
them. Being in the adjoint representation of the electroweak group, 
the Majorana mass term of
such triplets is gauge invariant. In terms of the usual and compact
two-by-two notation for triplets, the beyond the SM interactions are
described by the Lagrangian (with implicit flavour summation):
\begin{equation}
\label{Lfermtriptwobytwo}
{\cal L}=Tr [ \overline{\Sigma} i \slash \hspace{-2.5mm} D  \Sigma ] 
-\frac{1}{2} Tr [\overline{\Sigma}  M_\Sigma \Sigma^c 
                +\overline{\Sigma^c} M_\Sigma^* \Sigma] 
- \tilde{\phi}^\dagger \overline{\Sigma} \sqrt{2}Y_\Sigma L 
-  \overline{L}\sqrt{2} {Y_\Sigma}^\dagger  \Sigma \tilde{\phi}\, ,
\end{equation} 
with $L\equiv (l,\nu)^T$, $\phi\equiv (\phi^+, \phi^0)^T\equiv
(\phi^+, (v+H+i \eta)/\sqrt{2})^T$, $\tilde \phi = i \tau_{2} \phi^*$,
$\Sigma^c \equiv C \overline{\Sigma}^T$ and with, for each fermionic
triplet,
\begin{eqnarray}
\Sigma&=&
\left(
\begin{array}{ cc}
   \Sigma^0/\sqrt{2}  &   \Sigma^+ \\
     \Sigma^- &  -\Sigma^0/\sqrt{2} 
\end{array}
\right), \quad 
\Sigma^c=
\left(
\begin{array}{ cc}
   \Sigma^{0c}/\sqrt{2}  &   \Sigma^{-c} \\
     \Sigma^{+c} &  -\Sigma^{0c}/\sqrt{2} 
\end{array}
\right), \nonumber\\
D_\mu &=& \dv_\mu-i\sqrt{2} g \left(
\begin{array}{ cc}
   W^3_\mu/\sqrt{2}  &   W_\mu^+ \\
     W_\mu^- &  -W^3_\mu/\sqrt{2} 
\end{array}
\right)\,.
\end{eqnarray}
Without loss of generality, in the following we will assume that we
start from the basis where $M_\Sigma$ is real and diagonal.  In order
to consider the mixing of the triplets with the charged leptons, it is
convenient to express the four degrees of freedom of each charged
triplet in terms of a single Dirac spinor:
\begin{equation}
 \label{Psi}
\Psi\equiv\Sigma_R^{+ c} + \Sigma_R^-\,.
\end{equation}
The neutral fermionic triplet components on the other hand can be left
in two-component notation, since they have only two degrees of
freedom and mix with neutrinos, which are also described by
two-component fields. This leads to the Lagrangian
\begin{eqnarray}
\label{Lfull-ft-2}
{\cal L}&=& \overline{\Psi} i \dv \Psi  
+ \overline{\Sigma_R^0} i \dv  \Sigma^0_R  
-  \overline{\Psi}M_\Sigma \Psi -
        \left( \overline{\Sigma^{0}_R} \frac{{M_\Sigma}}{2}  \Sigma_R^{0c} \,+  \,\text{h.c.}\right) 
\nonumber \\
&+&g \left(W_\mu^+ \overline{\Sigma_R^0} \gamma_\mu  P_R\Psi 
 +  W_\mu^+ \overline{\Sigma_R^{0c}} \gamma_\mu  P_L\Psi   \,+  \,\text{h.c.}
 \right) - g\, W_\mu^3 \overline{\Psi} \gamma_\mu  \Psi 
 \nonumber\\
\nonumber \\ 
&-&  \left( \phi^0 \overline{\Sigma_R^0} Y_\Sigma \nu_{L}+ \sqrt{2}\phi^0 \overline{\Psi} Y_\Sigma l_{L}
+     \phi^+ \overline{\Sigma_R ^0} Y_\Sigma l_{L} - \sqrt{2}\phi^+
 \overline{{\nu_{L}}^c}
Y^{T}_\Sigma \Psi   \,+  \,\text{h.c.}\right)\,.
\end{eqnarray}
The mass term of the charged sector shows then the usual aspect for
Dirac particles:
\begin{equation}
{\cal L} \owns -(\overline{l_R}\,\, \overline{\Psi_R} )
\,\,
\left(
\begin{array}{ cc}
   m_l  &   0 \\
      {Y_\Sigma} v &  {M_\Sigma} 
\end{array}
\right) \,\,
\left(
\begin{array}{ c}
   l_L \\
  \Psi_L 
\end{array}
\right) \,\,- 
(\overline{l_L}\,\, \overline{\Psi_L} )
\,\,
\left(
\begin{array}{ cc}
   m_l  &   Y_\Sigma^\dagger v \\
     0 &  {M_\Sigma} 
\end{array}
\right) \,\,
\left(
\begin{array}{ c}
   l_R \\
  \Psi_R \,
\end{array}
\right)\, ,
\label{chargedfullmassmatrix}
\end{equation}
with $v\equiv \sqrt{2} \langle \phi^0 \rangle=246$~GeV.  The symmetric
mass matrix for the neutral states is on the other hand given by
\bea
{\cal L}& \owns& -(\overline{\nu_L}\,\, \overline{\Sigma^{0c}} )
\left(
\begin{array}{ cc}
  0  &   {Y_\Sigma}^\dagger v/2\sqrt{2} \\
   {Y_\Sigma}^* v/2\sqrt{2} &  {M_\Sigma}/2 
\end{array}
\right) 
\left(
\begin{array}{ c}
   \nu_L^c \\
   \Sigma^0 
\end{array}
\right) \, 
\nn\\
&&-(\overline{\nu_L^c}\,\, \overline{\Sigma^{0}} )\left(
\begin{array}{ cc}
  0  &   {Y_\Sigma}^T v/2\sqrt{2} \\
   {Y_\Sigma} v/2\sqrt{2} &  {M_\Sigma}/2 
\end{array}
\right) 
\left(
\begin{array}{ c}
   \nu_L \\
   \Sigma^{0c}
\end{array}
\right)\, .
\label{neutralfullmassmatrix}
\eea
%

\noindent{\bf Diagonalization of the mass matrices}

To calculate the $l_1 \rightarrow l_2 \gamma$ decay rates, we will
work in the mass eigenstates basis. As it happens with any Dirac mass,
the charged lepton mass matrix can be diagonalized by a bi-unitary
transformation
\begin{equation}
\left(
\begin{array}{ c}
   l_{L,R} \\
  \Psi_{L,R} 
\end{array}
\right) = U_{L,R}
\left(
\begin{array}{ c}
   l'_{L,R} \\
   \Psi'_{L,R}
\end{array}
\right)\,,
\end{equation}
where $U_{L,R}$ are $(3+n)$-by-$(3+n)$ matrices, if $n$ triplets are
present. On the contrary, the symmetric neutral lepton mass matrix can
be diagonalized by a single unitary matrix
\begin{equation}
\left(
\begin{array}{ c}
   \nu_{L} \\
   \Sigma^{0c}
\end{array}
\right) = U_0
\left(
\begin{array}{ c}
   \nu'_{L} \\
   \Sigma'^{0c}
\end{array}
\right).
\end{equation}
It is convenient to write the mixing matrices in terms of
three-leptons-plus-$n$-triplets sub-blocks
\begin{equation}
U_{L}\equiv
\left(
\begin{array}{ cc}
   U_{Lll} &   U_{L l\Psi} \\
     U_{L\Psi l} & U_{L \Psi\Psi}  
\end{array}
\right) \,,\,
U_{R}\equiv
\left(
\begin{array}{ cc}
   U_{Rll} &   U_{R l\Psi} \\
     U_{R\Psi l} & U_{R \Psi\Psi}  
\end{array}
\right) \,,\,
U_{0}\equiv 
\left(
\begin{array}{ cc}
   U_{0 \nu \nu } &   U_{0 \nu \Sigma} \\
     U_{0 \Sigma \nu} & U_{0 \Sigma \Sigma}  
\end{array}
\right)\,. 
\label{mixdef}
\end{equation}
In the following we will calculate the decay rates at ${\cal
O}((Y_{\Sigma}v/M_\Sigma)^2)$, which is a good approximation as long
as $M_\Sigma$ is sufficiently big compared to $Y_{\Sigma}v$.  In order
to do so it can be checked that it is enough to calculate all the
mixing matrix elements at order ${\cal O}([(Y_{\Sigma}v,
m_l)/M_\Sigma]^2)$. We obtain:
\be
\begin{array}{llll}
U_{Lll}=1-\epsilon & U_{L l\Psi}=Y_{\Sigma}^\dagger M^{-1}_\Sigma v &
U_{L \Psi l}=- M^{-1}_\Sigma Y_{\Sigma} v & U_{L\Psi\Psi} = 1-\epsilon' \\
U_{Rll}=1 & U_{R l\Psi}=m_l Y_{\Sigma}^{\dagger}M_{\Sigma}^{-2}  v &
U_{R \Psi l}=- M_{\Sigma}^{-2}Y_{\Sigma} m_l v & U_{R\Psi\Psi}=1 \\
U_{0\nu\nu}=(1-\frac{\epsilon}{2}) U_{PMNS} & 
U_{0 \nu \Sigma}= Y_{\Sigma}^\dagger M^{-1}_\Sigma \frac{v}{\sqrt{2}} &
U_{0\Sigma \nu}=-M^{-1}_\Sigma Y_{\Sigma} \frac{v}{\sqrt{2}} U_{0\nu\nu} &
U_{0 \Sigma \Sigma}=(1-\frac{\epsilon'}{2})
\label{mixingsU}
\end{array}
\ee
where $\epsilon=\frac{v^2}{2}Y_\Sigma^\dagger M^{-2}_\Sigma Y_\Sigma$,
$\epsilon'=\frac{v^2}{2}M^{-1}_\Sigma Y_\Sigma Y_\Sigma^\dagger
M^{-1}_\Sigma$ and $U_{PMNS}$ is the lowest order neutrino mixing
matrix which is unitary.  Note that $\epsilon$ is nothing but the
coefficient of the unique low energy dimension-six operator induced by
the triplets, once they have been integrated
out~\cite{ABBGH}.\footnote{The $\epsilon'$ contribution does not
appear in the low energy effective theory as it involves external
$\Sigma$'s.}  Eq.~(\ref{mixingsU}) shows as expected that the
$(3+n)$-by-$(3+n)$ mixing matrices $U_{L,R,0}$ are unitary but the
various submatrices are not.  The neutrino mass matrix in this model
is given by\footnote{As for the masses of the charged leptons, they
are essentially unaffected by the presence of the $\Sigma$'s as the
difference between the physical masses of the $l'$ and the ones of the
$l$'s, $m_l$, is of order $m_l Y_\Sigma^2 v^2/M_\Sigma^2$.}:
\begin{equation}
m_\nu=-\frac{v^2}{2} Y_\Sigma^T\frac{1}{M_\Sigma} Y_\Sigma\,.
\end{equation}
%

\noindent{\bf Lagrangian in the mass basis}

After the diagonalization of the mass matrices, we obtain the following
Lagrangian in the mass basis
(omitting from now on the primes on the
mass eigenstate fields):
\begin{equation}
\mathcal{L}=\mathcal{L}_{Kin}+\mathcal{L}_{CC}+\mathcal{L}_{NC}+\mathcal{L}_{H,\eta}+\mathcal{L}_{\phi^{-}}\, ,
\label{fulllagrangian}
\end{equation}
where
\bea
\label{CC}
\mathcal{L}_{CC}&=&\frac{g}{\sqrt{2}}\left(\begin{array}{cc}\overline{l} & \overline{\Psi}\end{array}\right)\gamma^{\mu}W^{-}_{\mu}\left(P_L g^{CC}_L+P_R g^{CC}_R\sqrt{2}\right)\left(\begin{array}{c}\nu \\ \Sigma\end{array}\right)+\textrm{h.c}.\\
%
\label{NC}
\mathcal{L}_{NC}&=&\frac{g}{cos\theta_W}\left(\begin{array}{cc}\overline{l} & \overline{\Psi}\end{array}\right)\gamma^{\mu}Z_{\mu}\left(P_L g^{NC}_L+P_R g^{NC}_R\right)\left(\begin{array}{c}l \\ \Psi\end{array}\right)\\
%
\label{Heta}
\mathcal{L}_{H,\eta}&=&\frac{g}{2M_W}\left(\begin{array}{cc}\overline{l} & \overline{\Psi}\end{array}\right)H\left(P_L g^{H}_L+P_R g^{H}_R\right)\left(\begin{array}{c}l \\ \Psi\end{array}\right)\nn\\
&+&i\frac{g}{2M_W}\left(\begin{array}{cc}\overline{l} & \overline{\Psi}\end{array}\right)\eta\left(P_L g^{\eta}_L+P_R g^{\eta}_R\right)\left(\begin{array}{c}l \\ \Psi\end{array}\right)\\
\label{phi}
\mathcal{L}_{\phi^{-}}&=&-\phi^{-}\overline{l}\frac{g}{\sqrt{2}M_W}\left\{\left(P_L g^{\phi^{-}}_{L_{\nu}}+P_R g^{\phi^{-}}_{R_{\nu}}\right)\nu+\left(P_L g^{\phi^{-}}_{L_{\Sigma}}+P_R g^{\phi^{-}}_{R_{\Sigma}}\right)\Sigma\right\}+\textrm{h.c.}
\eea
with
\bea
g^{CC}_{L}&=&\left(\begin{array}{cc}
g^{CC}_{L_{l\nu}}&g^{CC}_{L_{l\Sigma}}\\
g^{CC}_{L_{\Psi\nu}}&g^{CC}_{L_{\Psi\Sigma}}\end{array}\right)
= \left(\begin{array}{cc}
\left(1+\epsilon\right)U_{0_{\nu\nu}} & 
            -Y_{\Sigma}^{\dagger}M_{\Sigma}^{-1}\frac{v}{\sqrt{2}}\\
0 & \sqrt{2}\left(1-\frac{\epsilon'}{2}\right)\end{array}\right)\\
g^{CC}_{R}&=&\left(\begin{array}{cc}
g^{CC}_{R_{l\nu}}&g^{CC}_{R_{l\Sigma}}\\
g^{CC}_{R_{\Psi\nu}}&g^{CC}_{R_{\Psi\Sigma}}\end{array}\right)= 
\left(\begin{array}{cc}
0 & -m_lY_{\Sigma}^{\dagger}M_{\Sigma}^{-2}v\\
-M_{\Sigma}^{-1}Y^{*}_{\Sigma}U^{*}_{0_{\nu\nu}}\frac{v}{\sqrt{2}} & 
             1-\frac{\epsilon'^{*}}{2}\end{array}\right) \\
g^{NC}_{L}&=&\left(\begin{array}{cc}
g^{NC}_{L_{ll}}&g^{NC}_{L_{l\Psi}}\\
g^{NC}_{L_{\Psi l}}&g^{NC}_{L_{\Psi\Psi}}\end{array}\right)= 
\left(\begin{array}{cc}
\frac{1}{2}-cos^2\theta_W-\epsilon & 
      \frac{1}{2}Y_{\Sigma}^{\dagger}M_{\Sigma}^{-1}v\\
\frac{1}{2}M_{\Sigma}^{-1}Y_{\Sigma}v & 
      \epsilon'-cos^2\theta_W\end{array}\right)\\
g^{NC}_{R}&=&\left(\begin{array}{cc}
g^{NC}_{R_{ll}}&g^{NC}_{R_{l\Psi}}\\
g^{NC}_{R_{\Psi l}}&g^{NC}_{R_{\Psi\Psi}}\end{array}\right)= 
\left(\begin{array}{cc}
1-cos^2\theta_W & m_lY_{\Sigma}^{\dagger}M_{\Sigma}^{-2}v\\
 M_{\Sigma}^{-2}Y_{\Sigma}m_lv & -cos^2\theta_W \end{array}\right)\\
\label{ghl}
g^{H}_{L}&=&\left(\begin{array}{cc}
g^{H}_{L_{ll}}&g^{H}_{L_{l\Psi}}\\
g^{H}_{L_{\Psi l}}&g^{H}_{L_{\Psi\Psi}}\end{array}\right)= 
\left(\begin{array}{cc}
m_l\left(3\epsilon-1\right) & -m_lY_{\Sigma}^{\dagger}M_{\Sigma}^{-1}v\\
-Y_{\Sigma}v\left(1-\epsilon\right)-M_{\Sigma}^{-2}Y_{\Sigma}m_l^2v & 
                          \dots \end{array}\right)\\
g^{H}_{R}&=&\left(\begin{array}{cc}
g^{H}_{R_{ll}}&g^{H}_{R_{l\Psi}}\\
g^{H}_{R_{\Psi l}}&g^{H}_{R_{\Psi\Psi}}\end{array}\right)= 
\left(\begin{array}{cc}
\left(3\epsilon-1\right)m_l & 
-\left(1-\epsilon\right)Y_{\Sigma}^{\dagger}v-m_l^2Y_{\Sigma}^{\dagger}M_{\Sigma}^{-2}v\\
-M_{\Sigma}^{-1}Y_{\Sigma}m_lv & \dots \end{array}\right)\\
g^{\eta}_{R}&=&\left(\begin{array}{cc}
g^{\eta}_{R_{ll}}&g^{\eta}_{R_{l\Psi}}\\
g^{\eta}_{R_{\Psi l}}&g^{\eta}_{R_{\Psi\Psi}}\end{array}\right)= 
\left(\begin{array}{cc}
-\left(\epsilon+1\right)m_l & 
\left(1-\epsilon\right)Y_{\Sigma}^{\dagger}v-m_l^2Y_{\Sigma}^{\dagger}M_{\Sigma}^{-2}v\\
-M_{\Sigma}^{-1}Y_{\Sigma}m_lv & \dots \end{array}\right)\\
\label{getal}
g^{\eta}_{L}&=&\left(\begin{array}{cc}
g^{\eta}_{L_{ll}}&g^{\eta}_{L_{l\Psi}}\\
g^{\eta}_{L_{\Psi l}}&g^{\eta}_{L_{\Psi\Psi}}\end{array}\right)= 
\left(\begin{array}{cc}
m_l\left(\epsilon+1\right) & m_lY_{\Sigma}^{\dagger}M_{\Sigma}^{-1}v\\
-Y_{\Sigma}v\left(1-\epsilon\right)+M_{\Sigma}^{-2}Y_{\Sigma}m_l^2v & 
            \dots \end{array}\right)
\eea
and
\bea
\left\{\begin{array}{l}
g^{\phi^{-}}_{L_{\nu}}=m_l U_{0_{\nu\nu}}\\
g^{\phi^{-}}_{R_{\nu}}=-\left(1-\epsilon\right)m_{\nu}^{*} U^{*}_{0_{\nu\nu}}\end{array}\right. \quad
\left\{\begin{array}{l}
g^{\phi^{-}}_{L_{\Sigma}}=m_lY_{\Sigma}^{\dagger}M_{\Sigma}^{-1}\frac{v}{\sqrt{2}}\\
g^{\phi^{-}}_{R_{\Sigma}}=\left(1-\epsilon\right)Y_{\Sigma}^{\dagger}\frac{v}{\sqrt{2}}\left(1-\frac{\epsilon'^{*}}{2}\right)-\sqrt{2}m_{\nu}^{*}Y_{\Sigma}^{T}M_{\Sigma}^{-1}v\end{array}\right. \, .
\eea
The dots in Eqs.~(\ref{ghl})-(\ref{getal}) refer to $\Psi$-$\Psi$
interactions which we omit here since they do not contribute to the
one-loop $l_1 \rightarrow l_2 \gamma$ rates.

\section{$\boldsymbol{\mu \rightarrow e \gamma}$ and $\boldsymbol{\tau \rightarrow l \gamma}$ decays}

In the following we perform the calculation of the $\mu \rightarrow e
\gamma$ rate. The $\tau$ decay rates will be obtained
straightforwardly from it later on. As it is well-known, the on-shell
transition $\mu\rightarrow e \gamma$ is a magnetic transition so that
its amplitude can be written, in the $m_e\rightarrow 0$ limit, as :
\begin{eqnarray}
T\left(\mu\rightarrow e \gamma\right)=A\times\overline{u_e}\left(p-q\right)\left[iq^{\nu}\varepsilon^{\lambda}\sigma_{\lambda\nu}\left(1+\gamma_5\right)\right]u_{\mu}\left(p\right)\, ,
\end{eqnarray}
with $\varepsilon$ the polarization of the photon, $p_{\mu}$ the
momentum of the incoming muon, $q_{\mu}$ the momentum of the
outgoing photon and $\sigma_{\mu\nu}=\frac{i}{2}\left[\gamma_{\mu},\gamma_{\nu
}\right]$. Using the Gordon decomposition we can rewrite it as
\begin{eqnarray}
\label{gordonampli}
T\left(\mu\rightarrow e \gamma\right)=
A\times\overline{u_e}\left(p-q\right)\left(1+\gamma_5\right)
\left(2p\cdot\varepsilon-m_{\mu}\varepsilon\!\!\!/\right)u_{\mu}\left(p\right)\, .
\end{eqnarray}
In the following we will calculate only the $p\cdot\varepsilon$
terms. The terms proportional to $\varepsilon\!\!\!/$ can be recovered
from the $p\cdot\varepsilon$ terms through
Eq.~(\ref{gordonampli}). All in all, this gives:
\begin{equation}
\Gamma(\mu \rightarrow e \gamma)=\frac{m_\mu^3}{4 \pi} |A|^2\,.
\end{equation}
%

\begin{figure}[t]
\centering
\includegraphics{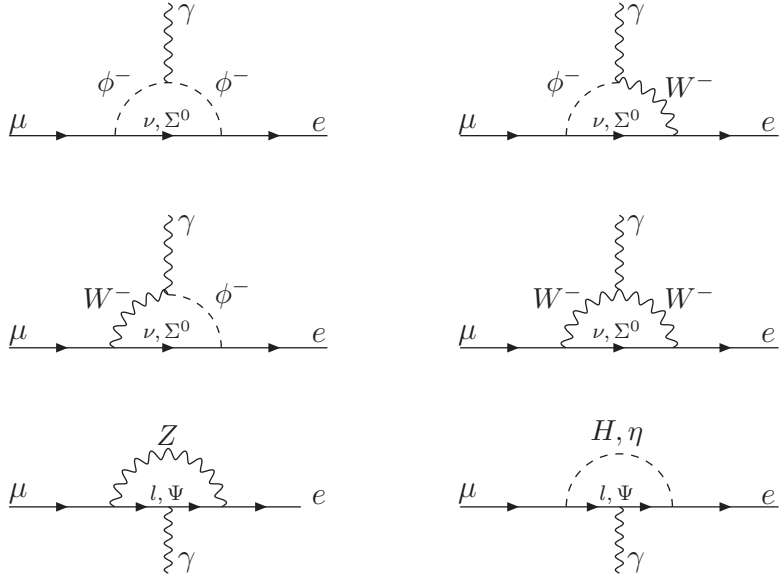}
\caption{Diagrams contributing to $\mu\rightarrow e \gamma$. $\phi^\pm,\, \eta$
are the three Goldstone boson associated with the $W^-$ and $Z$ bosons. $H$ stands for the
physical Higgs boson.}
\label{muegammatypeIII}
\end{figure}

\noindent{\bf $\boldsymbol{\mu \rightarrow e \gamma}$ amplitude and decay rate}

In the mass eigenstate basis, from the Lagrangian of
Eqs.~(\ref{CC})-(\ref{phi}), there are fourteen diagrams contributing
to $\mu \rightarrow e \gamma$, as shown in Fig.~1. The detailed
calculation is presented in the appendix~\footnote{ General formulae for radiative fermion decays have been derived in detail in Ref.~\cite{Lavoura}, although restricted to the case in which all fermion masses arise from the standard Higgs mechanism. In consequence, isospin invariant mass terms as those essential in seesaw models were not taken into account.}
. In the limit in which
$M_\Sigma\gg M_W$, at ${\cal O}((\frac{Y_\Sigma v}{M_{\Sigma}})^2)$,
the total amplitude is given by:
\begin{eqnarray}
\label{totalamplitude}
T\left(\mu\rightarrow e\gamma\right)&=&
i\frac{{G^{SM}_F}}{\sqrt{2}}\frac{e}{32\pi^2}m_{\mu}\overline{u_e}\left(p-q\right)\left(1+\gamma_5\right)i \sigma_{\lambda\nu} \varepsilon^{\lambda} q^\nu u_{\mu}\left(p\right)\nonumber\\
&\times& \left\{\left(\frac{13}{3}+C\right)\epsilon_{e\mu}-\sum_i x_{\nu_i}\left(U_{PMNS}\right)_{ei}\left(U_{PMNS}^\dagger\right)_{i\mu}\right\}\, ,
\end{eqnarray}
where $C=-6,56$ and $x_{\nu_{i}}\equiv \frac{m^2_{\nu_i}}{M_W^2}$.
Note that the second term is the usual contribution from neutrino
mixing~\cite{Bilenky:1987ty}, while the first one is the explicit
contribution of the fermion triplet(s). As well known, a GIM
cancellation operates in the second term. The total decay rate is then
given by:
\begin{eqnarray}
\Gamma\left(\mu\rightarrow e\gamma\right)=\frac{{G^{SM}_F}^2e^2m_{\mu}^5}{8192\pi^5}\left|\left(\frac{13}{3}+C\right)\epsilon_{e\mu}-\sum_i x_{\nu_i}\left(U_{PMNS}\right)_{ei}\left(U_{PMNS}^\dagger\right)_{i\mu}\right|^2
\end{eqnarray}
and the branching ratio reads
\begin{equation}
\label{brratio}
Br\left(\mu\rightarrow e\gamma\right)=\frac{3}{32}\frac{\alpha}{\pi}\left|\left(\frac{13}{3}+C\right)\epsilon_{e\mu}-\sum_i x_{\nu_i}\left(U_{PMNS}\right)_{ei}\left(U_{PMNS}^\dagger\right)_{i\mu}\right|^2\, .
\end{equation}
$\tau \rightarrow l \gamma$ decays can be obtained from
Eq.~(\ref{brratio}) by replacing $\mu$ by $\tau$, $e$ by $l$ and by
multiplying the obtained result by $Br(\tau \rightarrow e \nu_\tau
\bar{\nu}_e)=(17.84 \pm 0.05) \cdot 10^{-2}$~\cite{PDG}.

\section{Phenomenology}

From the result above, it is not surprising that in general we expect
a very tiny $\mu \rightarrow e \gamma$ rate. For instance, omitting
flavour indices, for a given value of $M_\Sigma$ we would expect in
general from the seesaw formula that $Y_{\Sigma}^2 \simeq m_\nu
M_\Sigma/v^2\sim \sqrt{\delta m^2_{atm}} M_\Sigma/v^2\sim
M_\Sigma/(10^{15}\,\hbox{GeV})$. This gives $\epsilon\sim
m_\nu/M_\Sigma\sim 10^{-25} \, (10^{15}\,\hbox{GeV}/M_\Sigma)$ and
$x_\nu\sim\delta m^2_{atm.}/M_W^2\sim 10^{-24}$ which leads to
$Br\left(\mu\rightarrow e\gamma\right)\sim 10^{-52}\cdot (10^{15}\,
\hbox{GeV}/M_\Sigma)^2$, far below the present upper limit $1.2 \cdot
10^{-11}$. In this case, even for $M_{\Sigma}$ as low as 100~GeV, we
get $Br\left(\mu\rightarrow e\gamma\right)\sim 10^{-26}$.  Similarly,
for $\tau \rightarrow \mu \gamma$ and $\tau \rightarrow e \gamma$, we
get both rates of order $ 10^{-53} (10^{15}\,\hbox{GeV}/M_\Sigma)^2$,
far below the present upper limit $4.5\cdot10^{-8}$ and
$1.1\cdot10^{-7}$, respectively.

There are cases, however, in which the branching ratio can be much
larger without any fine-tuning of the Yukawa couplings and mass
parameters. This is the case if neutrino masses are generated through
``direct lepton violation'' (DLV) (see Ref.~\cite{ABBGH}), i.e.~if
neutrino masses are directly proportional to a small lepton number
violating scale rather than inversely proportional to a high
scale. DLV appears naturally in the type-II seesaw model, since two
scales are present there: the mass of the heavy scalar triplet
$M_\Delta$ and the dimension-full trilinear coupling $\mu$ between the
scalar triplet and two Higgs doublets. In this case $m_\nu \sim
Y_\Delta \mu v^2/M_\Delta^2$, where $Y_\Delta$ is the Yukawa coupling,
but $Br(\mu \rightarrow e \gamma) \sim Y_\Delta^4
M_W^4/M_\Delta^4$. If the scale $\mu$ is sufficiently small to
suppress neutrino masses, $Y_\Delta/M_\Delta$ can be large enough to
generate visible effects in rare lepton decays. A similar pattern can
be realized also in the type-III seesaw, if besides a high scale
$M_\Sigma$, a low scale $\mu$, responsible for lepton number
violation, is present. This has indeed been studied in the context of
type-I seesaw~\cite{doubleseesaw,ABBGH}, but it can be applied here as
well. In this case the $\epsilon_{e\mu}$ term in
Eqs.~(\ref{totalamplitude})-(\ref{brratio}) is enhanced to much larger
values and the $x_{\nu_i}$ term can be neglected.

With such a pattern the $\mu \rightarrow e \gamma$ branching ratio
could be as large as $\sim10^{-4}$ for the extreme case where the
Yukawa couplings would be as large as unity with triplets as light as
few hundreds GeV.  This shows that the present experimental bound is
already relevant to exclude too large values of the Yukawas associated
to too small values of the triplet mass.  The present experimental
bounds on the branching ratios give the following constraints on the
$\epsilon_{\alpha\beta}$ 
coefficients:\footnote{Note that these bounds show that
the approximation we made in the above to work only at first order in
$Y^2 v^2/M_\Sigma^2$ is justified.}
\begin{eqnarray}
|\epsilon_{e \mu}|&=&
\frac{v^2}{2}\,|Y_\Sigma^\dagger\frac{1}{M_\Sigma^\dagger}\frac{1}{ M_\Sigma}Y_{\Sigma}|_{\mu e}\ \lesssim 1.1 \cdot 10^{-4}
\label{bound1C}\\
|\epsilon_{\mu \tau}|&=&
\frac{v^2}{2}\,|Y_\Sigma^\dagger\frac{1}{M_\Sigma^\dagger}\frac{1}{ M_\Sigma}Y_{\Sigma}|_{\tau \mu}\ \lesssim 1.5 \cdot 10^{-2}
\label{bound2C}\\
|\epsilon_{e \tau }|&=&
\frac{v^2}{2}\,|Y_\Sigma^\dagger\frac{1}{M_\Sigma^\dagger}\frac{1}{ M_\Sigma}Y_{\Sigma}|_{\tau e}\ \lesssim 2.4 \cdot 10^{-2}\, .
\label{bound3C}
\end{eqnarray}

\noindent
{\bf Comparison of $\boldsymbol{l \rightarrow l' \gamma}$ and 
$\boldsymbol{l \rightarrow 3 l'}$ decays}
  
The bounds of Eqs.~(\ref{bound1C})-(\ref{bound3C}) from $l \rightarrow
l' \gamma$ decays turn out to be on the same parameters $\epsilon$ as
the ones obtained from $\mu \rightarrow 3e$ or $\tau \rightarrow 3l$
decays, derived in Ref.~\cite{ABBGH}.  This can be understood from the
fact that, at order $1/M_\Sigma^2$, for example for $\mu \rightarrow e
\gamma$ and $\mu \rightarrow 3e$, there is only one way to combine two
Yukawa couplings and two inverse $M_\Sigma$ mass matrices to induce a
$\mu$-$e$ transition along a same fermionic line: through the
combination $\epsilon_{e \mu}$ (i.e.~the flavour structure of the
$\mu$-to-$e$ fermionic line is the same for both processes, it
corresponds to a $\mu$ which mixes with a fermion triplet which mixes
with an electron).  This can also be understood from the related fact
that the number of independent parameters contained in the
coefficients of the dimension five operators (proportional to the
neutrino mass matrix) and dimension six operators (encoded in the
$\epsilon_{\alpha \beta}$ \cite{ABBGH}) of the low energy theory
(obtained in the limit of large fermion triplet mass) equals the
number of independent parameters of the original theory. This implies
that any physical transition studied at order $1/M_\Sigma^2$,
necessarily has to be proportional to the dimension six operator
coefficients, and there is only one which gives a $\mu$-to-$e$
transition: $\epsilon_{e \mu}$.

As a result we obtain the following fixed ratios for these branching
ratios:
\begin{eqnarray}
Br(\mu \rightarrow e \gamma)&=&
1.3 \cdot 10^{-3} \cdot Br(\mu \rightarrow eee)\,, \label{relation1C}\\
Br(\tau \rightarrow \mu \gamma)&=&
1.3 \cdot 10^{-3} \cdot Br(\tau \rightarrow \mu \mu \mu)=
2.1 \cdot 10^{-3}\cdot Br(\tau^-\rightarrow e^- e^+\mu^-)\,,
\label{relation2C}\\
Br(\tau \rightarrow e \gamma)&=&
1.3 \cdot 10^{-3} \cdot Br(\tau \rightarrow e e e)\, \,=
2.1 \cdot 10^{-3}\cdot Br(\tau^-\rightarrow \mu^-\mu^+ e^-)\,.
\label{relation3C}
\end{eqnarray}
The ratios are much smaller than unity because $l \rightarrow 3l'$ is
induced at tree level through mixing of the charged leptons with the
charged components of the fermion triplets \cite{ABBGH}, while $l
\rightarrow l' \gamma$ is a one-loop process.  The results of
Eqs.~(\ref{relation1C})-(\ref{relation3C}) hold in the limit where
$M_\Sigma\gg M_{W,Z,H}$, as they are based on Eq.~(\ref{brratio}).
Not taking this limit, i.e.~using Eq.~(\ref{completetotalamplitude})
of the Appendix, for values of $M_\Sigma$ as low as $\sim 100$ GeV,
these ratios can vary around these values by up to one order of
magnitude. Numerically it turns out that the bounds in
Eqs.~(\ref{bound1C})-(\ref{bound3C}) are thus not as good as the ones
coming from $\mu \rightarrow eee$, $\tau \rightarrow eee$ and $\tau
\rightarrow \mu \mu \mu$ decays, which give $|\epsilon_{e \mu}| < 1.1
\cdot 10^{-6}$, $|\epsilon_{\mu \tau}| < 4.9 \cdot 10^{-4}$,
$|\epsilon_{e \tau}| < 5.1 \cdot 10^{-4}$ respectively (using the
experimental bounds: $Br(\mu \rightarrow eee) < 1 \cdot
10^{-12}$~\cite{PDG}, $Br(\tau \rightarrow eee) < 3.6 \cdot 10^{-8}$
\cite{Belle2} and $Br(\tau \rightarrow \mu \mu \mu) < 3.2 \cdot
10^{-8}$ \cite{Belle2}).\footnote{Note that these bounds from $\tau$
decays are better than the ones quoted in Table.~8 of
Ref.~\cite{ABBGH}, as we have used the new experimental limits on
$\tau \rightarrow 3l$ decays of Ref.~\cite{Belle2}.  
This also leads
to the new following bounds: $|\epsilon_{\mu \tau}| < 5.6 \cdot
10^{-4}$ (from $Br(\tau \rightarrow e^+ e^- \mu^-) < 2.7 \cdot
10^{-8}$) and $|\epsilon_{e \tau}| < 7.2 \cdot 10^{-4}$ (from $Br(\tau
\rightarrow \mu^+ \mu^- e^-) < 4.1 \cdot 10^{-8}$). 
We thank
M.~Nemev\v sek for pointing to us the existence of
Ref.~\cite{Belle2}.}  This shows that even if the upper limits on $\mu
\rightarrow e \gamma$ and $\tau \rightarrow l \gamma$ are improved in
the future by three or two orders of magnitude respectively, the $\mu
\rightarrow 3e$ and $\tau \rightarrow 3l$ will still provide the most
competitive bounds on the $\epsilon_{\alpha\beta}$ ($\alpha \neq
\beta)$.  This can be clearly seen from the bounds, $Br(\mu
\rightarrow e \gamma) < 10^{-15}$, $Br(\tau \rightarrow \mu \gamma) <
4 \cdot 10^{-11}$ and $Br(\tau \rightarrow e \gamma) < 5 \cdot
10^{-11}$, that one obtains from
Eqs.~(\ref{relation1C})-(\ref{relation3C}) using the experimental
bounds on the $l \rightarrow 3l'$ decays.

This leads to the conclusion that the observation of one leptonic
radiative decay by upcoming experiments would basically rule out the
seesaw mechanism with only triplets of fermions, i.e.~with no other
source of lepton flavour changing new physics. To our knowledge this
is a unique result.

This is different from other seesaw models. For instance, in type I
seesaw, for the same reasons as for the type-III model, 
the ratios of Eqs.~(\ref{relation1C})-(\ref{relation3C}) are
also fixed at order $1/M_N^2$, but unlike for this type-III model, 
both processes are instead realized at
one-loop. As a result, generically, $l \rightarrow l' \gamma$ dominates over $l
\rightarrow 3 l'$ because the latter suffers an extra $\alpha$
suppression. On the other hand, in type II seesaw, no definite
predictions for these ratios can be done, because both types of decays
depend on different combinations of the parameters~\cite{ABBGH}. This
stems from the fact that in the type-II model the Yukawa coupling
$Y_{\Delta}$ couples a scalar triplet to two light fermions, so it
carries two light lepton flavour indices, instead of one in the type-I
and type-III models. As a result there are several combinations of the
Yukawa couplings which can lead to a $\mu$-to-$e$ transition in this
model.\footnote{For instance the $\mu \rightarrow 3e$ transition
involves the combination $Y_{\Delta\mu e}Y^\dagger_{\Delta e e}$ while the
$\mu \rightarrow e \gamma$ involve the combination $Y_{\Delta\mu l}
Y^\dagger_{\Delta l e}$ with $l=e,\mu,\tau$ see e.g.~\cite{ABBGH}. }

\section{$\boldsymbol{\mu}$ to $\boldsymbol{e}$ conversion in atomic nuclei}

Beside $l\rightarrow l' \gamma$ and $ l\rightarrow 3 l$ decays,
fermion triplets can also induce $\mu$ to $e$ conversion in atomic nuclei. The relevant  
diagram turns out to be a tree level one, as for $l \rightarrow 3 l$ decays, where $\mu$ goes to $e + Z$ with the $Z$ connected to a $u$ or $d$ quark fermion line. For the reasons given above, or simply from the fact that this diagram involves exactly the same $\mu$-$e$-$Z$ vertex as the $\mu \rightarrow eee$ decay, $\mu$ to $e$ conversion gives a constraint on the same $\varepsilon_{e \mu}$ 
parameter than from $\mu \rightarrow eee$ decay (or than from $\mu \rightarrow e \gamma$ decay).
Using the experimental upper bound for the $\mu$ to $e$ conversion rate to total nucleon muon capture rate ratio for ${}^{48}_{22}Ti$ nuclei, $R^{\mu \rightarrow e} < 4.3 \cdot 10^{-12}$ \cite{sindrum}, the bound one obtains actually turns out to be even more stringent than from $\mu \rightarrow eee$:
\begin{equation}
| \varepsilon_{e \mu}| < 1.7 \times 10^{-7}
\end{equation}
This bound can be straightforwardly obtained by determining the quark-lepton
effective interaction induced by the $Z$ exchange
\bea
{\cal L}_{eff}&=&- \sqrt{2} G_F (\bar{l}_i \gamma^\alpha P_L g_{Lij}^{NC} l_j)
\times\nn\\
&\times&\left(\bar{u} \gamma_\alpha [(1-\frac{8}{3} \sin^2 \theta_W) -\gamma_5) ] u + \bar{d} \gamma_\alpha 
[ (-1+\frac{4}{3} \sin^2 \theta_W)+\gamma_5 ] d \right)
\eea
which using standard formula, for example Eq.~(2.16) of Ref.~\cite{bernartom}, gives
\begin{equation}
R^{\mu \rightarrow e}=1.4 \cdot 10^1 \cdot |\varepsilon_{e \mu}|^2.
\end{equation}
This leads to the following fixed ratio predictions for  ${}^{48}_{22}Ti$
\begin{eqnarray}
Br(\mu \rightarrow eee)=2.4 \cdot 10^{-1} R^{\mu \rightarrow e}\\
Br(\mu \rightarrow e \gamma)= 3.1 \cdot 10^{-4} R^{\mu \rightarrow e}
\end{eqnarray}
which allows further possibilities to test and/or exclude the model.
Results from the gold nuclei, which experimentally gives $R^{\mu \rightarrow e} < 7 \cdot 10^{-13}$ \cite{au}, are of same order of magnitude.
Note that the PRISM collaboration \cite{Prism} is expected to improve the experimental bound on $R^{\mu \rightarrow e}$ for the ${}^{48}_{22}Ti$ nuclei by several orders of magnitude in the long term.

\section{Summary}

We have calculated the $\mu \rightarrow e \gamma$ and $\tau
\rightarrow l \gamma$ decay rates in presence of one or more triplets
of fermions. As with right-handed neutrinos, the obtained rate is in
general extremely suppressed but in special cases (not necessarily
tuned) it can exceed the present experimental bounds.  Unlike for
other seesaw models, the observation of a leptonic radiative decay
rate close to the present bounds, would nevertheless be incompatible
with bounds which arise in this model from $l \rightarrow 3l'$
decays. Similarly it would be incompatible with the bound from $\mu$
to $e$ conversion we have determined.  This provides an interesting
possibility to exclude this model as the unique low energy source of
lepton flavour changing new physics.

\section*{Acknowledgments}
We acknowledge discussions with E.~Fern\'andez-Mart\'\i nez.  We especially thank Anna Rossi 
for having pointed out to us that the bound from $\mu$-$e$ conversion can actually be more stringent than the one from $\mu \rightarrow eee$ decay. The
authors received partial support from CICYT through the project
FPA2006-05423, as well as from the Comunidad Aut\'onoma de Madrid
through Proyecto HEPHACOS; P-ESP-00346.  T.H. thanks the FNRS-FRS for
support. A.A and F.B acknowledge the support of the Agence Nationale de
la Recherche ANR through the project JC05-43009-NEUPAC.

\section*{Appendix}

The fourteen diagrams of Fig.~1 can be grouped according to the
fermion circulating in the loop. Performing the calculation in the 't
Hooft-Feynman gauge, after loop integration, the various amplitudes,
at ${\cal O}((\frac{Y_\Sigma v}{M_{\Sigma}})^2)$, are:
\begin{eqnarray}
T^{\phi^-,W^-}_{\nu_i}&=&T^{\phi^-}_{\nu_{i}}+T^{\phi^-,W^-}_{\nu_{i}}+T^{W^-,\phi^-}_{\nu_{i}}+T^{W^-}_{\nu_{i}}=\nonumber\\
&=&i\frac{{G^{SM}_F}}{\sqrt{2}}\frac{e}{32\pi^2}m_{\mu}\overline{u_e}\left(p-q\right)\left(1+\gamma_5\right)\left(2p\cdot\varepsilon\right)u_{\mu}\left(p\right)\left[\left(U_{0_{\nu\nu}}\right)_{ei}\left(U_{0_{\nu\nu}}^\dagger\right)_{i\mu}F_1\left(x_{\nu_i}\right)\right.\nonumber\\
&&\left.+\left(\epsilon\,U_{0_{\nu\nu}}\right)_{ei}\left(U_{0_{\nu\nu}}^\dagger\right)_{i\mu}F_2\left(x_{\nu_i}\right)+\left(U_{0_{\nu\nu}}\right)_{ei}\left(U_{0_{\nu\nu}}^\dagger\epsilon\right)_{i\mu}F_3\left(x_{\nu_i}\right)\right]\\[.5cm]
%
T^{\phi^-,W^-}_{\Sigma_i}&=&
T^{\phi^-}_{\Sigma_{i}}+T^{\phi^-,W^-}_{\Sigma_{i}}
+T^{W^-,\phi^-}_{\Sigma_{i}}+T^{W^-}_{\Sigma_{i}}=\nonumber\\
&=&i\frac{{G^{SM}_F}}{\sqrt{2}}\frac{e}{32\pi^2}m_{\mu}
\overline{u_e}\left(p-q\right)\left(1+\gamma_5\right)
\left(2p\cdot\varepsilon\right)u_{\mu}\left(p\right)\nonumber\\
&&\bigg\{
\left(Y^{\dagger}_{\Sigma}M^{-1}_{\Sigma}\right)_{ei}
\left(M^{-1}_{\Sigma}Y_{\Sigma}\right)_{i\mu}\frac{v^2}{2}F_4(x_{\Sigma_i})
+\left(Y^{\dagger}_{\Sigma}M^{-1}_{\Sigma}\right)_{ei}
\left(M^{-1}_{\Sigma}Y_{\Sigma}\epsilon\right)_{i\mu}\frac{v^2}{2}x_{\Sigma_i}
F_5(x_{\Sigma_i})\nn\\
&&+\frac{1}{M_W^2}\bigg[\left(Y^{\dagger}_{\Sigma}\right)_{ei}
\left(\epsilon'^{T}Y_{\Sigma}\right)_{i\mu}\frac{v^2}{4}
+\left(Y^{\dagger}_{\Sigma}\right)_{ei}\left(M^{-1}_{\Sigma}Y_{\Sigma}^* 
m_{\nu}^T\right)_{i\mu}v^2\bigg]F_5(x_{\Sigma_i})\nn\\
&&+\frac{1}{M_W^2}\bigg[ \left(Y^{\dagger}_{\Sigma}\epsilon'^*\right)_{ei}
\left(Y_{\Sigma}\right)_{i\mu}\frac{v^2}{4}
+\left( m^*_\nu Y_\Sigma^T M^{-1}_\Sigma\right)_{ei}
\left(Y_\Sigma\right)_{i\mu} v^2\bigg]F_6(x_{\Sigma_i})\nn\\
&&+\left(\epsilon Y^\dagger_\Sigma
M^{-1}_\Sigma\right)_{ei}\left( M^{-1}_\Sigma Y_\Sigma \right)_{i\mu}
\frac{v^2}{2}x_{\Sigma_i}F_6(x_{\Sigma_i})
\bigg\}
\eea
\begin{eqnarray}
T^{Z,H,\eta}_{\Psi_i}&=&
T^{Z}_{\Psi_i}+T^{H}_{\Psi_i}+T^{\eta}_{\Psi_i}=\nonumber\\
&=&i\frac{{G^{SM}_F}}{\sqrt{2}}\frac{e}{32\pi^2}m_{\mu}
\overline{u_e}\left(p-q\right)\left(1+\gamma_5\right)
\left(2p\cdot\varepsilon\right)u_{\mu}\left(p\right)\nonumber\\
&&\bigg[\left(Y^{\dagger}_{\Sigma}M^{-1}_{\Sigma}\right)_{ei}
\left(M^{-1}_{\Sigma}Y_{\Sigma}\right)_{i\mu}\frac{v^2}{2}
\left(F_7\left(y_{\Sigma_i}\right)+F_8\left(z_{\Sigma_i}\right)\right)\nn\\
&&-\left(\epsilon Y^{\dagger}_{\Sigma}M^{-1}_{\Sigma}\right)_{ei}
\left(M^{-1}_{\Sigma}Y_{\Sigma}\right)_{i\mu}\frac{v^2}{2}
\left(F_8\left(y_{\Sigma_i}\right)+F_8\left(z_{\Sigma_i}\right)\right)\nn\\
&&-\left(Y^{\dagger}_{\Sigma}M^{-1}_{\Sigma}\right)_{ei}
\left(M^{-1}_{\Sigma}Y_{\Sigma}\epsilon\right)_{i\mu}\frac{v^2}{2}
\left(F_9\left(y_{\Sigma_i}\right)+F_9\left(z_{\Sigma_i}\right)\right)
\bigg]\\[.5cm]
%
T^{Z,H,\eta}_{l_i}&=&
T^{Z}_{l_i}+T^{H}_{l_i}+T^{\eta}_{l_i}=\nonumber\\
&=&i\frac{{G^{SM}_F}}{\sqrt{2}}\frac{e}{32\pi^2}m_{\mu}
\overline{u_e}\left(p-q\right)\left(1+\gamma_5\right)
\left(2p\cdot\varepsilon\right)u_{\mu}\left(p\right)
\epsilon_{e\mu}{G}\left(y_{li},z_{l_i}\right)\, ,
\end{eqnarray}
where $x_{\nu_{i}}\equiv \frac{m^2_{\nu_i}}{M_W^2}$,
$x_{\Sigma_{i}}\equiv \frac{m^2_{\Sigma_i}}{M_W^2}$,
$y_{l_{i}}=\frac{m^2_{l_{i}}}{M_Z^2}$,
$z_{l_{i}}=\frac{m^2_{l_{i}}}{M_H^2}$,
$y_{\Sigma_{i}}=\frac{m^2_{\Sigma_i}}{M_Z^2}$,
$z_{\Sigma_i}=\frac{m^2_{\Sigma_i}}{M_H^2}$ and $F_i(x)$ and $G(x)$
are the following functions:
\begin{eqnarray}
\label{F1}
F_1(x)&=&\frac{10-43x+78x^2-49x^3+4x^4+18x^3\log(x)}{3(-1+ x)^4}\\
F_2(x)&=&\frac{2(5-24x+39x^2-20x^3+6x^2(-1+2x)\log(x))}{3(-1+x)^4}\\
F_3(x)&=&\frac{7-33x+57x^2-31x^3+6x^2(-1+3x)\log(x)}{3(-1+x)^4}\\
F_4(x)&=&\frac{-38+185x-246x^2+107x^3-8x^4+18(4-3x)x^2\log(x)}{3(-1+x)^4}\\
F_5(x)&=&\frac{1-6x+3x^2+2x^3-6x^2\log(x)}{3(-1+x)^4}\\
F_6(x)&=&\frac{7-12x-3x^2+8x^3-6x(-2+3x)\log(x)}{3(-1+x)^4}\\
F_7(x)&=&\frac{40-46x-3x^2+2x^3+7x^4+18x(4-3x)\log(x)}{3(-1+x)^4}\\
F_8(x)&=&\frac{x(-16+45x-36x^2+7x^3+6(-2+3x)\log(x))}{3(-1+x)^4}\\
F_9(x)&=&\frac{x(2+3x-6x^2+x^3+6x\log(x))}{3(-1+x)^4}
\eea
\bea
G(y_{l_i},z_{l_i})&=\,\delta_{ie}&
\left[8\left(\frac{1}{2}-cos^2\theta_W\right)
\frac{4-9y_{l_i}+5y_{l_i}^3+6(1-2y_{l_i})y_{l_i}\log(y_{l_i})}{6(-1+y_{l_i})^4}\right]\nn\\
&+\,\delta_{i\mu}&
\bigg[z_{l_i}\frac{16-45z_{\l_i}+36z_{\l_i}^2-7z_{\l_i}^3-6(-2+3z_{\l_i})\log(z_{\l_i})}{2(-1+z_{\l_i})^4}\nonumber\\
&&+8\left(\frac{1}{2}-cos^2\theta_W\right)
\frac{4-9y_{l_i}+5y_{l_i}^3+6(1-2y_{l_i})y_{l_i}\log(y_{l_i})}{6(-1+y_{l_i})^4}\nonumber\\
&&-8\left(1-cos^2\theta_W\right)
\frac{2(-1+y_{l_i}^2-2y_{l_i}\log(y_{l_i}))}{(-1+y_{l_i})^3}\nonumber\\
&&-y_{l_i}\frac{-20+39y_{l_i}-24y_{l_i}^2+5y_{l_i}^3+6(-2+y_{l_i})\log(y_{l_i})}{6(-1+y_{l_i})^4}\bigg]\, .
\end{eqnarray}
Since $y_{l_i},\ z_{l_i},\ x_{\nu_{i}} \ll 1$, it is a
good approximation to take the lepton flavour conserving quantities
$y_{l_i}$ and $z_{l_i}$ to zero and to keep only the linear term in the
flavour changing quantities $x_{\nu_{i}}$:
\begin{eqnarray}
F_1(x_{\nu_{i}})&\simeq& \frac{10}{3}-x_{\nu_i}\\
F_2(x_{\nu_{i}})&\simeq& \frac{10}{3}-\frac{8}{3}x_{\nu_{i}} \\
F_3(x_{\nu_{i}})&\simeq& \frac{7}{3}-\frac{5}{3}x_{\nu_{i}}\\
G(y_i,z_i)&=& C=-6,56 \, .
\end{eqnarray}
Summing over $i$ and neglecting terms of ${\cal O}((Y_\Sigma
v/M_{\Sigma})^n)$ with $n>2$, we obtain:
\begin{eqnarray}
\label{amplitude1}
T^{\phi^-,W^-}_{\nu}&=&\sum_i T^{\phi^-,W^-}_{\nu_i}=\nn\\
&=&i\frac{{G^{SM}_F}}{\sqrt{2}}\frac{e}{32\pi^2}m_{\mu}
\overline{u_e}\left(p-q\right)\left(1+\gamma_5\right)
\left(2p\cdot\varepsilon\right)u_{\mu}\left(p\right)\nn\\
&&\bigg\{\frac{7}{3}\epsilon_{e\mu}-\sum_i
x_{\nu_i}\left(U_{PMNS}\right)_{ei}
\left(U_{PMNS}^\dagger\right)_{i\mu}\bigg\}\\
%
\label{amplitude2}
T^{\phi^-,W^-}_{\Sigma}&=&\sum_i T^{\phi^-,W^-}_{\Sigma_i}=\nn\\
&=&i\frac{{G^{SM}_F}}{\sqrt{2}}\frac{e}{32\pi^2}m_{\mu}\overline{u_e}
\left(p-q\right)\left(1+\gamma_5\right)\left(2p\cdot\varepsilon\right)
u_{\mu}\left(p\right)\nn\\
&&\bigg\{-\frac{8}{3}\epsilon_{e\mu}+
\sum_i \frac{v^2}{2} \left(Y_\Sigma^\dagger M_\Sigma^{-1}\right)_{ei}
\left(M_\Sigma^{-1} Y_\Sigma\right)_{i\mu} A(x_{\Sigma_i})
\bigg\}\\
\label{amplitude3}
T^{Z,H,\eta}_{l}&=&\sum_i T^{Z,H,\eta}_{l_i}=\nn\\
&=&i\frac{{G^{SM}_F}}{\sqrt{2}}\frac{e}{32\pi^2}m_{\mu}
\overline{u_e}\left(p-q\right)\left(1+\gamma_5\right)
\left(2p\cdot\varepsilon\right)u_{\mu}\left(p\right)\epsilon_{e\mu}
\times C
\eea
\bea
\label{amplitude4}
T^{Z,H,\eta}_{\Psi}&=&\sum_i T^{Z,H,\eta}_{\Psi_i}=\nn\\
&=&i\frac{{G^{SM}_F}}{\sqrt{2}}\frac{e}{32\pi^2}m_{\mu}
\overline{u_e}\left(p-q\right)\left(1+\gamma_5\right)
\left(2p\cdot\varepsilon\right)u_{\mu}\left(p\right)\nn\\
&&\bigg\{\frac{14}{3}\epsilon_{e\mu}+
\sum_i \frac{v^2}{2} \left(Y_\Sigma^\dagger M_\Sigma^{-1}\right)_{ei}
\left(M_\Sigma^{-1} Y_\Sigma\right)_{i\mu}
\Big(B(y_{\Sigma_i})+C(z_{\Sigma_i})\Big)\bigg\}\, ,
\end{eqnarray}
where
\bea
A(x_{\Sigma_i})&=&\frac{-30+153x_{\Sigma_i}-198x^2_{\Sigma_i}+75x^3_{\Sigma_i}
+18(4-3x_{\Sigma_i})x^2_{\Sigma_i}\log x_{\Sigma_i}}{3(x_{\Sigma_i}-1)^4}\\
B(y_{\Sigma_i})&=&\frac{33-18y_{\Sigma_i}-45y_{\Sigma_i}^2+30y_{\Sigma_i}^3+
18(4-3y_{\Sigma_i})y_{\Sigma_i}\log y_{\Sigma_i}}{3(y_{\Sigma_i}-1)^4}\\
C(z_{\Sigma_i})&=&\frac{-7+12z_{\Sigma_i}+3z_{\Sigma_i}^2-8z_{\Sigma_i}^3
+6(3z_{\Sigma_i}-2)z_{\Sigma_i}\log z_{\Sigma_i}}{3(z_{\Sigma_i}-1)^4}\, .
\eea
The total amplitude is then:
\begin{eqnarray}
\label{completetotalamplitude}
T\left(\mu\rightarrow e\gamma\right)&=&
i\frac{{G^{SM}_F}}{\sqrt{2}}\frac{e}{32\pi^2}m_{\mu}
\overline{u_e}\left(p-q\right)\left(1+\gamma_5\right)
\left(2p\cdot\varepsilon\right) u_{\mu}\left(p\right)\\
&\times& \Bigg\{\left(\frac{13}{3}+C\right)\epsilon_{e\mu}-\sum_i x_{\nu_i}
\left(U_{PMNS}\right)_{ei}
\left(U_{PMNS}^\dagger\right)_{i\mu}+\nn\\
&&\sum_i \frac{v^2}{2} \left(Y_\Sigma^\dagger M_\Sigma^{-1}\right)_{ei}
\left(M_\Sigma^{-1} Y_\Sigma\right)_{i\mu}
\Big(A(x_{\Sigma_i})+B(y_{\Sigma_i})+C(z_{\Sigma_i})\Big)\Bigg\}\nn\, .
\end{eqnarray}
This result is valid at ${\cal O}((\frac{Y_\Sigma v}{M_{\Sigma}})^2)$.
For $x_{\Sigma_i},y_{\Sigma_i},z_{\Sigma_i}\gg 1$, the additional
limit $x_{\Sigma_i},y_{\Sigma_i},z_{\Sigma_i}\rightarrow\infty$ can be
taken, which leads to the result displayed in the text,
Eq.~(\ref{totalamplitude}).


\end{document}